# Time-resolved energy transfer from single chloride-terminated nanocrystals to graphene

O. A. Ajayi[1,2]*, N. C. Anderson[3], M. Cotlet[4], N. Petrone[2], T. Gu[1], A. Wolcott[3], F. Gesuele[1], J. Hone[2], J. S. Owen[3], and C. W. Wong[1,2]*

[1]*Optical Nanostructures Laboratory, Center for Integrated Science and Engineering, Solid-State Science and Engineering, Columbia University, New York, New York, 10027, USA*

[2]*Department of Mechanical Engineering, Columbia University, New York, New York, 10027, USA*

[3] *Department of Chemistry, Columbia University, New York, New York, 10027, USA*

[4]*Brookhaven National Laboratory, Upton, New York, New York, 11973, USA*

We examine the time-resolved resonance energy transfer of excitons from single *n*-butyl amine-bound, chloride-terminated nanocrystals to two-dimensional graphene through time-correlated single photon counting. The radiative biexponential lifetime kinetics and blinking statistics of the individual surface-modified nanocrystal elucidate the non-radiative decay channels. Blinking modification as well as a 4× reduction in spontaneous emission were observed with the short chloride and *n*-butylamine ligands, probing the energy transfer pathways for the development of graphene-nanocrystal nanophotonic devices.

*Author email address: oaa2114@columbia.edu, cww2104@columbia.edu

Recent advances in ligand exchange techniques have paved the way for improved surface passivation of quantum dots.[1-3] In these techniques, long insulating organic ligands are replaced by short inorganic ligands. Recently, anionic halide ligands have been used to replace carboxylate ligands that balance the charge of the metal-rich nanoparticle surfaces, thus achieving charge-neutral nonstoichiometric particles without long chain organic surfactants.[4] These short compact halide ligands have been shown to effectively passivate mid-gap trap states in quantum dot films thus improving the open circuit voltage, leading to record power conversion efficiencies (8%).[5] Additionally, these short halide ligands are favorable for studies of the rich field of distant-dependent near field interactions.

Of particular interest are interactions between halide terminated quantum dots and two-dimensional materials,[6] which experience strongly enhanced coupling due to planar confinement and follow a $d^{-4}$ distance scaling[7-9] at short distances where electron-hole pair interactions dominate. Since the isolation of single layer graphene,[10] there has been an explosion of research of purely two-dimensional materials due to the unique physics anticipated at this atomic scale.[11-17] Graphene, with its broadband transparency[18] and high carrier mobility[19] provides a robust platform to explore light-matter interactions with photoexcited halide-terminated quantum dots. **Additionally, graphene's mechanical flexibility encourages its application in the emerging class of flexible electronics.** In the presence of graphene, a donor-acceptor interaction occurs providing competing pathways for photoluminescence through Förster-like resonant energy and/or charge transfer. Spectral overlap, electronic coupling, and proximity determine the kinetics of energy and charge transfer between the excited quantum dot and graphene. Resonant energy transfer occurs via dipole coupling to the two-dimensional plane of graphene with a $z^{-4}$ distance rate dependence.[6,19-21] Electronic excitation energy is transferred from the photoexcited dipole via Coulomb coupling to the graphene acceptor. Charge transfer is possible when suitable alignment occurs between the energy levels of the donor quantum dot and graphene acceptor.

Here we study the near-field interactions between quantum dots passivated by non-insulating, inorganic ligands and graphene, which allows us to probe **relaxation pathways relevant** to nanophotonic devices such as photodetectors. We measure the radiative lifetime and blinking statistics of single chloride-terminated CdSe nanocrystals on large-area chemical-vapor-deposited graphene to quantify the energy transfer rate and efficiency. A 4× reduction in the radiative lifetimes of the photoexcited single halide-terminated nanocrystals was observed on graphene clad glass substrates versus bare glass, corresponding to an energy transfer rate of $2.37 \times 10^8$ $s^{-1}$.

Chloride-terminated nanocrystals were synthesized by cleaving the native long carboxylate ligands on core-only 3.3 nm CdSe nanocrystals as detailed in ref 4. The resulting nanocrystals are passivated with chlorine ligands which balance the charge of the metal rich core and *n*-butylamine ligands (0.6 nm in length), which maintain the solubility of the nanocrystals. The short distance (~ 2.3 nm) from the *n*-butylamine bound nanocrystal center to graphene facilitates greater dipole energy transfer compared to ~ 4.5 nm from conventional oleic-acid capped nanocrystals of the same core diameter.

We probe the near field interactions between chloride-terminated nanocrystals and graphene by measuring the temporally resolved fluorescence of the nanocrystals on graphene-clad glass substrates and bare glass substrates. The comparative quenching of the fluorescence on graphene provides a measure of the non-radiative decay rate as will be discussed. Figure 2 compares the photoluminescence of individual chloride-terminated nanocrystals on graphene-clad glass and bare glass substrates, both pumped at 2 μW averaged power with 90 ps pulses at

440 nm. On bare glass, single nanocrystals were more easily discernible with ~ 170 counts per millisecond, while on graphene, the photon counts from the single nanocrystals are lower, ~ 100 counts per millisecond. Photon collection at the individual or few nanocrystal level allows us to gain insight into the exciton radiative and non-radiative relaxation rates[22-23] which can be masked by ensemble inhomogeneous broadening. As shown by the circled points of interest in Figure 2, individual nanocrystals were selected for lifetime measurements.

Individual nanocrystals are known to exhibit "blinking" behavior[24-26] with distinct "on" and "off" photoluminescence intensity states that can be characterized by probability distributions of their *on*-times [$P(t_{on})$] and *off*-times [$P(t_{off})$]. The probability distributions are modeled as power laws[27], with the off-times described by $P(t_{off}) = A t_{off}^{-m_{off}}$ and the on-times described by a truncated power law, $P(t_{on}) = A t_{on}^{-m_{on}} e^{-t_{on}/\tau_{on}}$. Figure 3 shows representative photoluminescence intensity traces of the individual chloride-terminated nanocrystals on comparative graphene and glass substrates, with a $\Delta t = 10$ ms binning window.

We confirm the presence of individual nanocrystals using power law statistics. The characteristic exponents for the off- and on-time distributions are extracted using a log-log least squares fitting method. For the off-time distribution, $m_{off}$ is found from the linear least-squares fitting of log[$P(t_{off})$] versus log[$t_{off}$] and for the on-time distribution, $t_{on}$ (saturation time) and $m_{on}$ are found from the least squares fitting to log[$P(t_{on})$] versus log[$t_{on}$]. Our results in Figure 3b and 3c show that the character of these individual chloride surface-modified nanocrystals (on glass) are in rough agreement with the $\tau_{on}/\tau_{off}$ exponents of conventional spherical nanocrystals at this binning time.[28] We also observe a smaller $\tau_{on}$, which is expected because $\tau_{on}$ decreases with the square of the applied power:[29] here we utilize a stronger pump power (2 µW) due to the low quantum yield (~ 3%) of these modified quantum dots at nanomolar concentration of amine. The chloride-terminated nanocrystals on chemical-vapor-deposited (CVD) graphene demonstrated different blinking statistics as compared to glass as shown in Figures 3f and 3g. Characteristic blinking is not observed on CVD graphene as is evident from the *poor* log-log fit to the expected power law distribution for the on ($P(t_{on}) = At_{on}^{-m_{on}} e^{-t_{on}/\tau_{on}}$) and off ($P(t_{off}) = At_{off}^{-m_{off}}$) times. Instead, varied fluctuations in intensity are observed in our measurements, as exampled in Figure 3d and 3e. The causes of the fluctuations are not completely understood, but we believe that the surface states of the chloride-terminated nanocrystals play a role in varied fluctuations.

To rigorously quantify the energy transfer and quenching (especially in the presence of blinking), we examined the radiative lifetime of the individual surface-modified nanocrystals on graphene. The energy transfer rate, $k_{et}$, and quenching can be experimentally determined from $k_{et} = \frac{1}{\tau'_{obs}} - \frac{1}{\tau_{obs}}$, where $\tau_{obs}$ is the radiative lifetime of the dipole on glass and $\tau'_{obs}$ is the radiative lifetime on graphene. The quenching (energy transfer rate) is of the form of Fermi's Golden Rule, where de-excitation of the photoexcited nanocrystal and subsequent electronic excitation of graphene is driven by Coulomb dipole interactions of the nanocrystal and graphene. Several theoretical models have been developed to account for the near-field interaction of graphene and an excited emitter, confirming a $1/z^4$ distance dependence where $z$ is the mean distance between the dipole center and the graphene surface. As with absorption[30], the fine structure constant has been shown to be central to a universal scaling law of energy transfer in graphene,[7] yielding a simple and powerful relation. We adapt this scaling to yield a calculated energy transfer rate, $k_{cal}$, described by:

$$k_{cal} \approx \frac{1}{\tau_o} \left[ \frac{9\nu\alpha}{256\pi^3 (E+1)^2} \left(\frac{\lambda_0}{z}\right)^4 \right] \tag{1}$$

where $\nu$ is a dipole orientation constant, $\nu = 1$ and $\nu = 2$ for parallel and perpendicular dipole orientation respectively, $\lambda_0$ is the free-space emission wavelength of the emitter, $E$ is the permittivity of the bulk medium (glass in this case), $\alpha$ is the fine structure constant, and $\tau_o$ is the radiative lifetime of the emitter in vacuum.

As shown in Figure 4, a biexponential lifetime decay was observed for individual chloride-terminated nanocrystals on both glass and graphene. The biexponential lifetime decay of these individual nanocrystals on glass can be attributed to fine splitting of the optical transitions in CdSe nanocrystal due to stochastic ground state dipole moments. These dipole moments arise from imperfect surface passivation.[31] With these chlorine-terminated quantum dots, imperfect surface passivation is in part due to the tendency of the amine ligands to dissociate from the nanocrystal surface at low ligand concentration in solution[32] and also to their slow evaporation from the nanocrystal surface when the nanocrystals are deposited on the substrate. The fast lifetime component ($\tau_1$) of the lifetime decay is attributed to recombination of the populated core-state (surface states) while the slow lifetime component ($\tau_2$) is attributed to exciton recombination.[33]

On graphene, both $\tau_1$ and $\tau_2$ components of the biexponential lifetime are shortened, as summarized in Figure 4b. This enhanced lifetime quenching is indicative of energy transfer of the excitons into graphene that is competitive with the present relaxation pathways: surface trap and exciton recombination.[33] The reduction of nanocrystal blinking on graphene further confirms that the energy transfer rate exceeds that of the competing pathway of photo-induced electron trapping rate of the blinking state.[6,34] With an amplitude-weighted average of the $\tau_1$ and $\tau_2$ lifetime components (given by $\tau = \sum_{i=1}^{n} \frac{A_i \tau_i}{\sum A_i}$), we observed an average lifetime $\tau$ of 13.2 ns (σ: 7.2 ns) on glass and 3.2 ns (σ: 2.2 ns) on graphene. This is also illustrated in Figure 4 and is a 4× reduction in the spontaneous emission lifetime. We note that the average $\tau_1$ on glass and graphene are 2.03 ns (σ: 0.50 ns) and 1.42 ns (σ: 0.28 ns) respectively. The average $\tau_2$ on glass and graphene are 30.49 ns (σ: 6.37 ns) and 25.31 ns (σ: 11.46 ns) respectively. With the amplitude-weighted lifetime, we obtain an experimentally observed resonant energy transfer rate of 2.4 ×10$^8$ s$^{-1}$. This yields an energy transfer efficiency *($\eta = 1-\tau/\tau_o$)* of 76%.

For direct comparison, we also investigated the energy transfer between similarly sized CdSSe/ZnS nanocrystals (emission 578 nm) onto graphene as shown in the insert of Figure 4. We expect the longer length of the oleic acid ligands terminating the nanocrystals to reduce energy transfer. We observe a 2.5× reduction in the spontaneous emission lifetime of these oleic acid capped nanocrystals (average lifetime on glass: 13.87 ns, σ: 10.79 ns and average lifetime on graphene: 5.46 ns, σ: 2.6 ns), corresponding to an energy transfer rate of 1.11 × 10$^8$ s$^{-1}$. A biexponential decay is also observed with the oleate-terminated nanocrystals. Here the average $\tau_1$ on glass and graphene are 6.14 ns (σ: 5.75 ns) and 3.11 ns (σ: 0.92 ns) respectively. The average $\tau_2$ on glass and graphene are 33.01 ns (σ: 6.13 ns) and 31.76 ns (σ: 8.77 ns) respectively.

We note that the calculated rate of energy transfer (2.4× 10$^8$ s$^{-1}$) from chloride-terminated nanocrystals onto graphene is on par with dipole-dipole transfer in nanocrystal assemblies[35] and nanocrystals on graphene-derived surfaces[36,37], albeit lower than intensity-derived energy transfer measurements of well-passivated core-shell nanocrystals on graphene[6] and recent molecule-graphene studies.[7] The ideal theoretical resonant energy transfer rate (from equation 1) is 4.0 ×10$^{10}$ s$^{-1}$ for the parallel dipole orientation ($k_{//}$) and 8.1 × 10$^{10}$ s$^{-1}$ for the perpendicular dipole orientation ($k_\perp$), yielding a dipole-averaged [k = (2/3)$k_{//}$ + (1/3)$k_\perp$] Förster energy transfer rate (1/$\tau_o$) of 5.4 × 10$^{10}$ s$^{-1}$. Considering the additional distance added from the presence of residual poly-methyl-methacrylate (PMMA: ~ 1 nm thickness), on the graphene surface, the theoretical resonant energy is adjusted to $1.28 \times 10^{10}$ s$^{-1}$. The energy transfer efficiency can be improved with thorough removal of any residual films such as the PMMA film on graphene, reduction of impurities at the substrate-graphene interface through a dry transfer technique[38], removal of the remaining alklyamines groups on the modified nanocrystal surface, and further passivation of the chloride-surface modified nanocrystals.

Our results have demonstrated significant near-field coupling between isolated surface-treated nanocrystals and larger-area monolayer CVD graphene. This study shows that considerable energy transfer rates (2.37 x10$^8$ s$^{-1}$) and energy transfer efficiencies (76%) are possible with chloride-terminated CdSe nanocrystals. Halide ligands are attractive for energy transfer applications because of their short length, however a trade-off exists between efficient coulombic coupling and surface stability. The high theoretical energy transfer rate greatly motivates further studies to optimize the ligand length towards improved energy transfer performance. Following these studies, numerous research areas are ripe for exploration with this system including plasmonics[39] and photo-detection.[40]

The authors acknowledge discussions with Zheyuan Chen and thank Nicolas Bais, Justin Abramson, Edward Judokusumo, Lance Kam, and Michael Sheetz, on the early single nanocrystal photoluminescence measurements. This work is supported primarily as part of the Center for Redefining Photovoltaic Efficiency through Molecule Scale Control, an Energy Frontier Research Center funded by the U.S. Department of Energy, Office of Science, Basic Energy Sciences under Award DE-SC0001085. The EFRC is supported by the New York State Office of Science, Technology, and Academic Research (NYSTAR) under contract C090147. O.A.A. acknowledges a Graduate Research Fellowship through the National Science Foundation. N.C.A. was supported by a National Science Foundation Graduate Research Fellowship under

Grant No. DGE07-07425. Research carried out in part at the Center for Functional Nanomaterials, Brookhaven National Laboratory, which is supported by the U.S. Department of Energy, Office of Basic Energy Sciences, under Contract No. DE-AC02-98CH10886.

**Figures:**

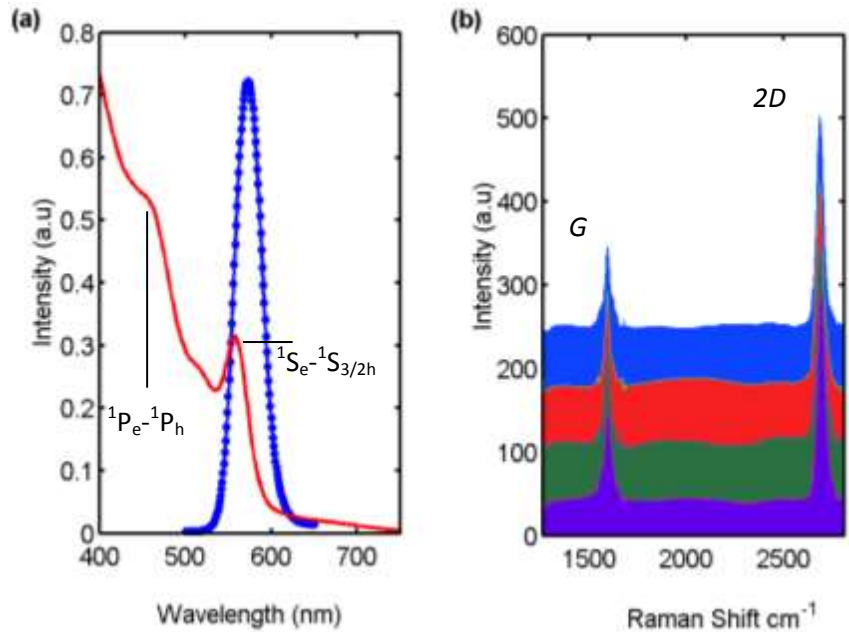

**Figure 1 |** Spectroscopic characterization of the chloride-terminated nanocrystals and graphene. **a,** Photoluminescence emission for the chloride-terminated nanocrystals, with peak intensity at 573 nm and first excitonic transition at 558 nm. **b,** Raman spectra of CVD graphene on glass, for four different spatial positions superimposed. The *G* and *2D* optical phonon bands homogeneity (centered at $1593 \pm 2.2$ nm and $2689 \pm 3.9$ nm respectively), single *G* peak linewidth symmetry, and linewidths (*G* band full-width half-maximum of $30.75 \pm 4.9$ nm and *2D* band full-width half-maximum of $44.65 \pm 2.5$ nm) indicate single layer graphene with good uniformity.

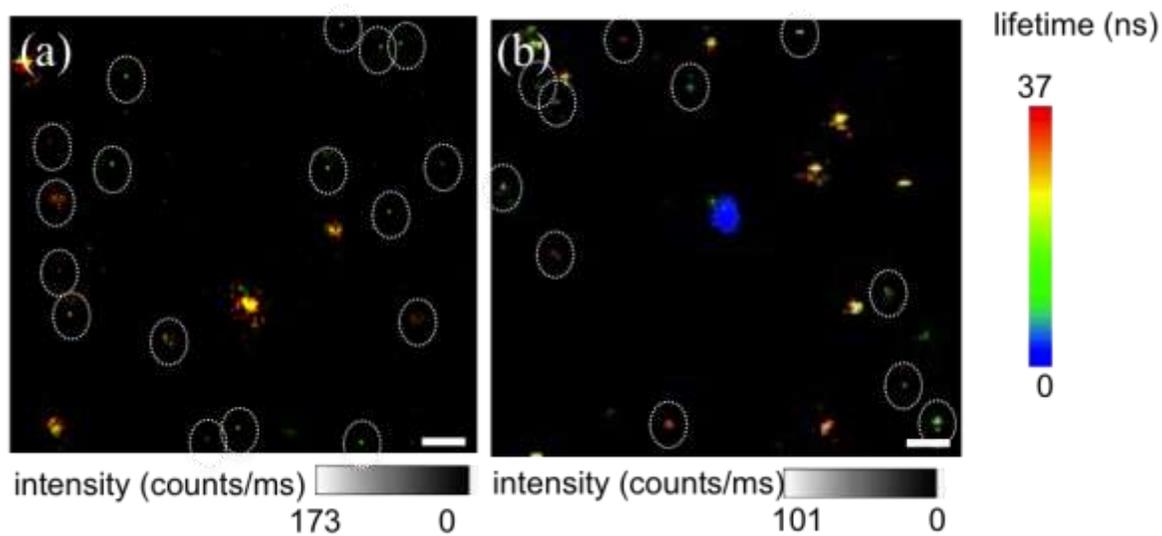

**Figure 2** | Single chloride-terminated nonstoichiometric nanocrystal photoluminescence spectroscopy **a,** Single chloride-terminated nanocrystal on glass, pump excitation 2 μW **b,** Single chloride-terminated nanocrystal on graphene, pump excitation 2 μW. The single nanocrystal photon counts are higher on glass (~ 170 counts per millisecond) versus on graphene (~ 100 counts per millisecond). Scale bar: 500 nm.

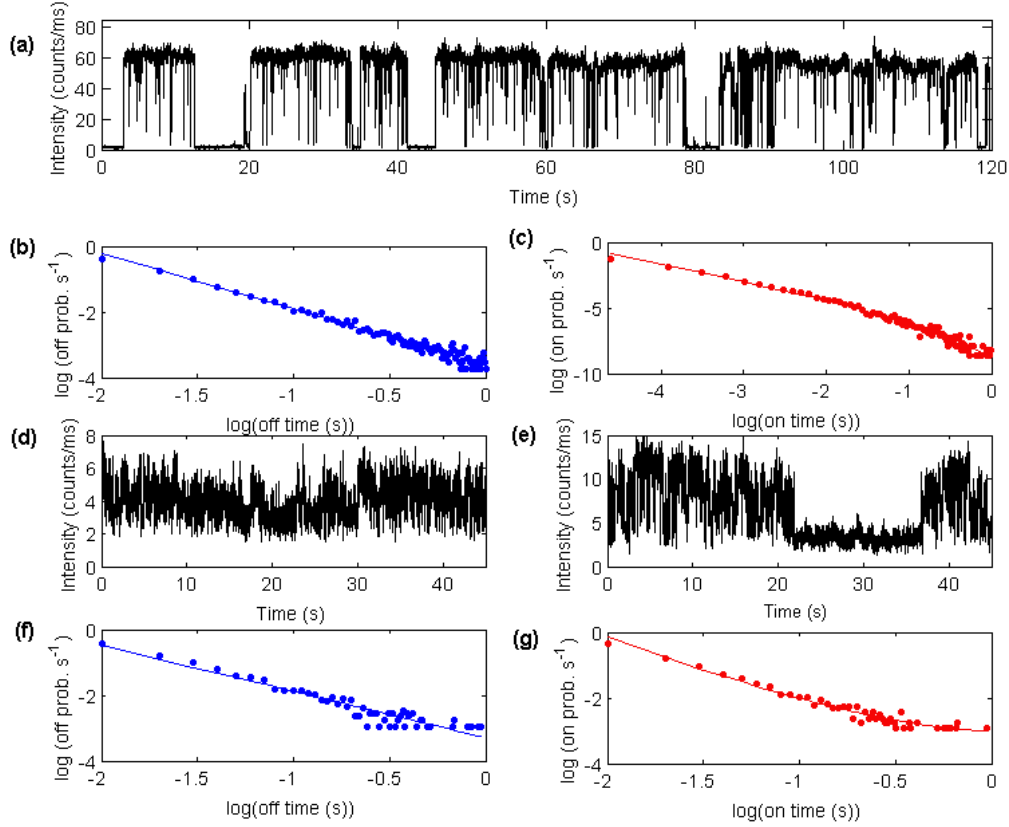

**Figure 3** | Blinking statistics of single chloride-terminated quantum dot on glass and graphene **a**, Representative intensity time trace of nanocrystal on glass. **b**, Off-time probability distribution (1/s) on glass $P(t_{off}) = At_{off}^{-m_{off}}$, $m_{off}$ = 1.7. **c**, On-time probability distribution on glass $P(t_{on}) = At_{on}^{-m_{on}}e^{-t_{on}/\tau_{on}}$, $\tau_{on}$ = 1.28, $m_{on}$ = 0.5. **d** and **e**, Intensity time traces of nanocrystal on graphene. **f**, Off-time probability distribution on graphene $P(t_{off}) = At_{off}^{-m_{off}}$, $m_{off}$ = 1.4. **g**, On-time probability distribution on graphene $P(t_{on}) = At_{on}^{-m_{on}}e^{-t_{on}/\tau_{on}}$, $\tau_{on}$ = 0.48 and $m_{on}$ = 2.3.

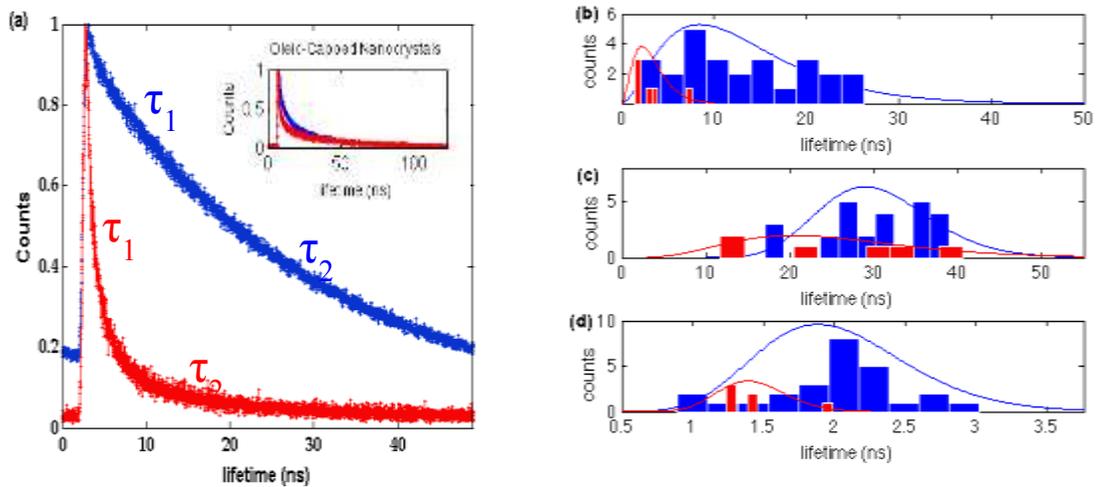

**Figure 4** | Time-resolved lifetime measurements of single chloride-terminated nanocrystal on glass (blue) and graphene (red). (b-d) Distribution of lifetime components fit to Gamma distribution **a,** Representative lifetime traces of single chlorine-terminated nanocrystal on glass and graphene. Insert: representative lifetime traces of single oleic-capped nanocrystals on glass and graphene **b,** Distribution of weighted lifetime on glass and graphene (chlorine-terminated nanocrystals). **c,** Distribution of long lifetime component ($\tau_2$) on glass and graphene (chlorine-terminated nanocrystals). **d,** Distribution of short lifetime component ($\tau_1$) on glass and graphene (chlorine-terminated nanocrystals).

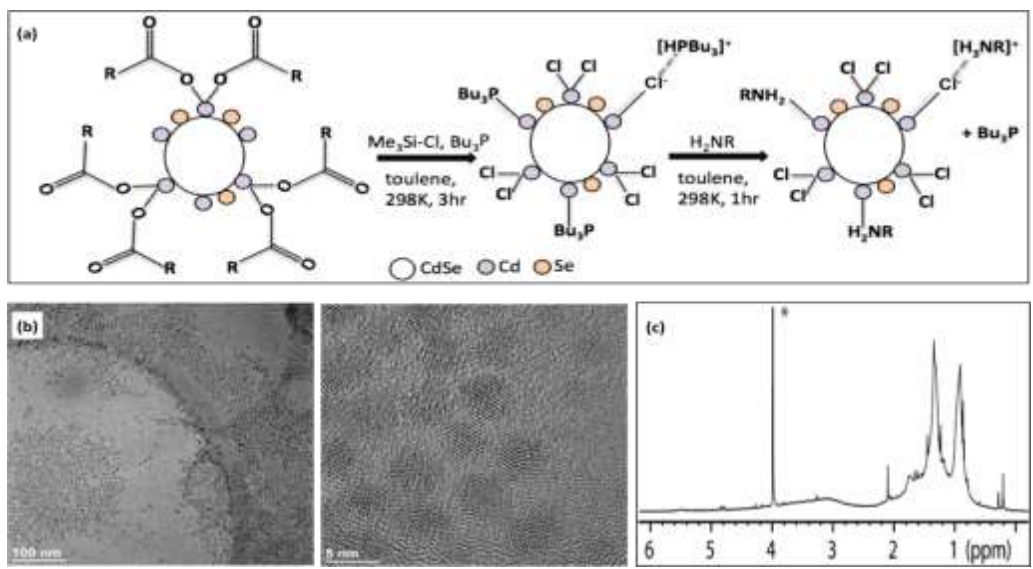

**Scheme 1**. Chemical synthesis and characterization of chloride-terminated CdSe nanocrystals. **a,** Schematic depiction of nanocrystal synthesis (Ref 4). **b,** High-resolution transmission electron micrographs of CdSe-CdCl$_2$/RNH$_2$ nanocrystals. Left scale bar: 50 nm. Right scale bar: 5 nm. **c,** $^1$H Nuclear magnetic resonance spectrum of CdSe-CdCl$_2$/RNH$_2$ with ferrocene standard.